\documentclass[aps,preprint,epsfig,rotate]{revtex4}
\begin{document}
%\begin{doublespace}
\title{Analytical formula for the Uehling potential}

 \author{Alexei M. Frolov}
 \email[E--mail address: ]{afrolov@uwo.ca}

 \author{David M. Wardlaw}
 \email[E--mail address: ]{dwardlaw@uwo.ca}

\affiliation{Department of Chemistry\\
 University of Western Ontario, London, Ontario N6H 5B7, Canada}

\date{\today}

\begin{abstract}

The closed analytical expression for the Uehling potential is derived. 
The Uehling potential describes the lowest-order correction on vacuum 
polarisation in atomic and muon-atomic systems. We also derive the 
analytical formula for the interaction potential between two 
electrically charged point particles which includes correction to the
vacuum polarisation, but has correct asymptotic behaviour at larger $r$.
Our three-term analytical formula for the Uehling potential opens a new 
avenue in the study of the vacuum polarisation in light atomic systems. 

\end{abstract}

\maketitle
\newpage

The effect of vacuum polarisation in light atoms and ions has attracted 
significant theoretical attention, since the early days of Quantum 
Electrodynamics \cite{Dir} - \cite{Paul} (see also discussion and references 
in \cite{PRA1976}). In 1935 Uehling \cite{Uehl} has produced the following 
formula for the interaction potential between two electric charges ($Q e$ 
and $e$) which contains an additional term responsible for the electric 
polarisation of the vacuum 
\begin{equation}
 \phi(r) = \frac{Q e}{r} \Bigl[ 1 + \frac{2 \alpha}{3 \pi} \cdot
 \int_1^{+\infty} exp(-2 \alpha^{-1} \xi r) \Bigl(1 + \frac{1}{2 \xi^2}
 \Bigr) \frac{\sqrt{\xi^2 - 1}}{\xi^2} d\xi \Bigr] = \frac{Q}{r} + U(r)
 \label{Int0}
\end{equation}
where we have used the atomic units $\hbar = 1, m_e = 1$ and $e = 1$. Here 
and henceforth $\alpha = \frac{e^2}{\hbar c}$ is the fine structure 
constant ($\alpha^{-1} \approx 137.03599911$, see, e.g., \cite{NIST}). The 
additional potential $U(r)$ in Eq.(\ref{Int0}) is the Uehling potential 
which is generated by the lowest order radiative correction to the 
electrodynamic Green's function, or photon Green's function, for short (for 
more details, see, e.g., \cite{AB}, \cite{Grei}, \cite{Fro1}). As follows 
from Eq.(\ref{Int0}) in light atoms the Uehling potential differs from zero 
only for very short atomic distances $r \le \Lambda_e$, where $\Lambda_e = 
\frac{\hbar}{m_e c} = \alpha a_0 \approx 3.861592 \cdot 10^{-11}$ $cm$ is the 
Compton wavelength of the electron and $a_0$ is the Bohr radius.

In numerous books and textbooks on Quantum Electrodynamics (see, e.g.,
\cite{AB}, \cite{Grei}) one can find a statement that the integral in
Eq.(\ref{Int0}) `cannot be evaluated in closed form but is easily solved
numerically'. In other words, it is widely assumed that the analytical
expression for the $U(r)$ potential does not exist. In order to show that
this statement is incorrect we proceed to obtain the closed analytical form
for the Uehling potential $U(r)$. First, consider the following integral
\begin{eqnarray}
 I_U(a) = \int_1^{+\infty} exp(-a \xi) \Bigl(1 + \frac{1}{2 \xi^2} \Bigr)
 \frac{\sqrt{\xi^2 - 1}}{\xi^2} d\xi \; \; \; , \label{Int1}
\end{eqnarray}
where $a$ is some positive numerical parameter. By using the substitution
$\xi = \cosh x$ we reduce $I_U$ to the form
\begin{eqnarray}
 I_U = \int_0^{+\infty} exp(-a \cosh x) \Bigl(1 - \frac{1}{2 \cosh^2 x}
 - \frac{1}{2 \cosh^4 x} \Bigr) dx = K_0(a) - \frac12 Ki_2(a) -
 \frac12 Ki_4(a) \; \; \; , \label{int2}
\end{eqnarray}
where $K_0(a)$ is the modified Bessel function of zero order (see, e.g, 
\cite{GR}, \cite{AS}, \cite{Treat}), i.e.
\begin{eqnarray}
 K_0(z) = \int_0^{\infty} exp(-z \cosh t) dt = \sum_{k=0}^{\infty} 
 (\psi(k+1) + \ln 2 - \ln z) \frac{z^{2 k}}{2^{2 k} (k!)^2} \; \; \; , 
 \label{macd}
\nonumber
\end{eqnarray}
where $\psi(k)$ is the Euler $psi$-function defined by Eq.(8.362) from
\cite{GR}. The functions $Ki_2(z)$ and $Ki_4(z)$ in Eq.(\ref{int2}) are the
recursive integrals of the $K_0(z) \equiv K_0(z)$ function, i.e.
\begin{eqnarray}
 Ki_1(z) = \int_z^{\infty} Ki_0(z) dz \; \; \; , \; \; \; and \; \; \;
 Ki_n(z) = \int_z^{\infty} Ki_{n-1}(z) dz  \; \; \; , \label{repin}
\end{eqnarray}
where $n \ge 1$. By using the relation given by Eq.(11.2.14) from \cite{AS} 
for the recursive integrals $Ki_n(z)$ (where $n \ge 0$) one can reduce 
Eq.(\ref{int2}) to the following form
\begin{eqnarray}
 I_U(z) = (1 + \frac{z^2}{12}) K_0(z) - \frac{z}{12} Ki_1(z) -
 (\frac56 + \frac{z^2}{12}) Ki_2(z) \; \; \; .
\end{eqnarray}
In the present case we have $z = a = 2 b r$ and the explicit formula for Uehling 
potential $U(r) = U(2 b r)$ takes the form
\begin{eqnarray}
 U(2 b r) = \frac{2 \alpha Q}{3 \pi r} \Bigl[ \Bigl(1 + \frac{b^2
 r^2}{3}\Bigr) K_0(2 b r) - \frac{b r}{6} Ki_1(2 b r) - \Bigl(\frac{b^2
 r^2}{3} + \frac{5}{6}\Bigr) Ki_2(2 b r) \Bigr] \; \; \; , \label{Ueh}
\end{eqnarray}
where the notation $Q$ stands for the electric charge of the nucleus and $b = 
\alpha^{-1}$. Now, Eq.(\ref{Int0}) takes the form
\begin{eqnarray}
 \phi(r) = \frac{Q}{r} \Bigl\{ 1 + \frac{2 \alpha}{3 \pi} \cdot
 \Bigl[ \Bigl(1 + \frac{r^2}{3 \alpha^2}\Bigr) 
 K_0\Bigl(\frac{2 r}{\alpha}\Bigr) - 
 \frac{r}{6 \alpha} Ki_1\Bigl(\frac{2 r}{\alpha}\Bigr)
 - \Bigl(\frac{5}{6} + \frac{r^2}{3 \alpha^2}\Bigr) 
 Ki_2\Bigl(\frac{2 r}{\alpha}\Bigr) \Bigr] \Bigr\} \nonumber \\
 = \frac{Q}{r} + U(r) \label{UehA}
\end{eqnarray}
In atomic units this potential coincides with the Coulomb interaction energy 
between the nucleus (electric charge is $Q e$) and the electron $e$. The potential
Eq.(\ref{UehA}) includes the lowest-order correction to the vacuum polarisation. 
As expected in atomic units this expression depends only upon the fine structure 
constant, i.e. upon the universal constants $\hbar$ and $c$, and electric charge 
$e$ (and $Q e$). The electron mass $m_e$ is also included in this formula, since 
the dimensionles variable $r$ in Eq.(\ref{UehA}) is $r = \frac{r}{a_0} = \frac{r 
m_e e^2}{\hbar^2}$, where $a_0$ is the Bohr radius. For muonic-helium atoms 
\cite{Grein2}, \cite{Plun} one needs to replace here $r \rightarrow r 
\Bigl(\frac{m_{\mu}}{m_e}\Bigr)$. In this study we do not want to discuss the 
vacuum polarisation in muonic atoms and ions. Note only that the generalisation of 
these formulae to the case of two interacting electric charges $q_1$ and $q_2$ is 
obvious and simple (see below). 

By using the known formulae for the limiting forms of the $K_0(2 b r), 
Ki_1(2 b r)$ and $Ki_2(2 b r)$ functions (see, e.g., \cite{GR}, \cite{AS} and 
\cite{Treat}) one finds the asymptotics of the Uehling potential which correspond 
to the cases when $r \ll \alpha a_0 = \Lambda_e$ and $r \gg \alpha a_0 = \Lambda_e$, 
respectively. In particular, the short range asymptotic of the $\phi(r) = 
\frac{Q}{r} + U(r)$ potential, Eq.(\ref{UehA}), takes the form (in atomic units)
\begin{equation}
 \phi(r)\mid_{r \rightarrow 0} \simeq \frac{Q}{r} \Bigl\{ 1 + \frac{\alpha}{3 
 \pi} \Bigl[ -\frac53 - 2 \gamma + 2 \ln \alpha - 2 \ln r \Bigr] \Bigr\} 
\end{equation}
where $\gamma \approx 0.5772156649\ldots$ is the Euler constant (see, e.g.,
\cite{GR}) and $\ln \alpha \approx -4.92024365857$. The long-range asymptotics
of the potential $\phi(r)$, Eq.(\ref{UehA}), is (in atomic units) 
\begin{equation}
 \phi(r)\mid_{r \rightarrow \infty} \simeq \frac{Q}{r} \Bigl\{ 1 + 
 \frac{\alpha^{\frac52}}{4 \sqrt{\pi} r^{\frac32}} \exp(-\frac{2}{\alpha} r) 
 \Bigr\} 
\end{equation}
Note that the long-range asymptotics of $\phi(r)$ decreases with $r$ exponentially. 
This is not correct physically, since the exponential function vanishes at $r 
\rightarrow +\infty$ very rapidly. In reality, such incorrect asymptotics of the 
interaction potential $\phi(r)$ at large distances $r$ are overweighted by the lowest 
order QED correction to the electromagnetic field $({\bf E}, {\bf H})$, or to the 
pure electric field ${\bf E}$ in our case, where ${\bf H} = 0$. Such a correction 
provides the correct power-type dependence of the interparticle potential at large 
disctances. It is directly related to the non-linearity of the Maxwell equation for 
the EM-field \cite{AB}, \cite{WK}.

The lowest order QED correction to the electric field is described by the Wichmann-Kroll 
potential $W_K(r)$ (in atomic units) \cite{WK}, \cite{Fro2}
\begin{equation}
  W_K(r) = -\frac{2 Q^3 \alpha^{7}}{225 \pi r^5} \nonumber
\end{equation}
Note that this potential is not regular at the origin. However, as follows from 
the Appendix of \cite{Fro2} such a singularity is formal, since the quibic equation 
(see Eq.(7) from \cite{Fro2}) which is used in the derivation of $W_K(r)$ is nor correct at
$r = 0$. The singularity at $r \rightarrow 0$ can be removed with the use of the substitution 
$r \rightarrow r + \alpha$ in atomic units, or $r \rightarrow r + \alpha a_0$ in regular units. 
It removes the singularity of the $W_K(r)$ potential at $r = 0$ and changes its behaviour at 
short interparticle distances $r \le \alpha$. However, at such distances the Wichmann-Kroll 
correction is not important and its contribution is significantly smaller than contribution 
from the Uehling potential. On the other hand, at large distances, e.g., for $r \ge 10 
\alpha a_0$ in light atoms, the contribution from the Wichmann-Kroll potential plays a 
leading role. Finally, we can write the correct expression for the Wichmann-Kroll potential 
$W_K(r)$ 
\begin{equation}
 \psi(r) = W_K(r) = -\frac{2 Q^3 \alpha^{7}}{225 \pi (r + \alpha)^5} \label{WK}
\end{equation}
This expression can directly be used in calculations of light atoms and ions. 

It is clear that the potential $W_K(r)$, Eq.(\ref{WK}), is always negative. In general, at $r 
\approx a_0$ this potential is very small in its absolute value. However, its overall 
contribution rapidly increases with the nuclear charge $Q$. Furthermore, it decreases with 
the distance $r$ as $\sim r^{-5}$, i.e. non-exponentially. It is clear that at large 
distances the Wichmann-Kroll correction will always exceed the contribution form the Uehling 
potential $U(r)$, Eq.(\ref{UehA}). Finally, we can say that the following interaction potential 
(in atomic units) of the two point electric charges $Qe$ and $e$
\begin{eqnarray}
 \Phi(r) = \frac{Q}{r} + U(r) + W_K(r) = \frac{Q}{r} + 
 \frac{2 Q \alpha}{3 \pi r} \cdot \Bigl[ \Bigl(1 + \frac{r^2}{3 \alpha^2}\Bigr) 
 K_0\Bigl(\frac{2 r}{\alpha}\Bigr) - \frac{r}{6 \alpha} 
 Ki_1\Bigl(\frac{2 r}{\alpha}\Bigr) \nonumber \\
 - \Bigl(\frac{5}{6} + \frac{r^2}{3 \alpha^2}\Bigr) 
 Ki_2\Bigl(\frac{2 r}{\alpha}\Bigr) \Bigr] 
 -\frac{2 Q^3 \alpha^{7}}{225 \pi (r + \alpha)^5} \label{total}
\end{eqnarray}
has the correct asymptotic behaviour both at small and large interparticle 
distances. In the case of the interaction between two point electric charges $q_1 
e$ and $q_2 e$ we need to replace in Eq.(\ref{total}) the factor $Q$ by the 
product $q_1 q_2$. The potential $\Phi(r_{12})$ takes the form
\begin{eqnarray}
 \Phi(r_{12}) = \frac{q_1 q_2}{r_{12}} + 
 \frac{2 q_1 q_2 \alpha}{3 \pi r_{12}} \cdot \Bigl[ \Bigl(1 + \frac{r_{12}^2}{3 
 \alpha^2}\Bigr) K_0\Bigl(\frac{2 r_{12}}{\alpha}\Bigr) - \frac{r_{12}}{6 \alpha} 
 Ki_1\Bigl(\frac{2 r_{12}}{\alpha}\Bigr) \nonumber \\
 - \Bigl(\frac{5}{6} + \frac{r_{12}^2}{3 \alpha^2}\Bigr) 
 Ki_2\Bigl(\frac{2 r_{12}}{\alpha}\Bigr) \Bigr] 
 -\frac{2 (q_1 q_2)^3 \alpha^{7}}{225 \pi (r_{12} + \alpha)^5} \label{totalt}
\end{eqnarray}
where $r_{12}$ is the distance between particles 1 and 2. 

Thus, we have derived the closed analytical expression for the Uehling potential
which represents the lowest order vacuum polarisation correction(s) in atomic
systems.  It is shown that the Uehling potential can be represented as a sum of 
the modified Bessel function of the zero order $K_0(2 b r)$ and the two recursive 
integrals of this function, i.e. the $K_1(2 b r)$ and $K_2(2 b r)$ functions 
(see Eq.(\ref{Ueh})). Based on this formula for the Uehling potential we also 
derived the general formula for the interaction between two point electric
charges which includes the lowest order QED corrections to the Coulomb potential.
Note that our formulae, Eqs.(\ref{total}) - (\ref{totalt}), provide the correct 
asymptotic behaviour at arbitrary interparticle distances. The formulae, 
Eqs.(\ref{total}) - (\ref{totalt}), can directly be used in highly accurate 
computations of the correction to the vacuum polarisation for the bound state 
energies in few-electron atoms and ions. By using the formula, Eq.(\ref{total}),
one can determine the electric field strength ${\bf E}$, i.e. the spatial gradient 
of the $\Phi(r_{12})$ potential, Eq.(\ref{totalt}). Such computations are 
straightforward, but the Fourier resolution of the corresponding `electrostatic' 
field takes a very complex form. Note also that for the potential $\Phi(r)$ given 
by Eq.(\ref{total}) the condition $\Delta \Phi(r) = 4 \pi Q \delta({\bf r})$, where 
$\Delta = \nabla \cdot \nabla = div(grad ...)$, is obeyed only approximately (in the
lowest order approximation upon $\alpha$).

Now, by taking into account our explicit and relatively simple three-term formula, 
Eq.(\ref{Ueh}), we can reject a widely spread conclusion that the Uehling potential 
cannot be represented by the closed analytical formula. Moreover, we have found 
that such a simple analytical formula drastically simlifies theoretical analysis of 
vacuum polarization in light atoms and ions. In our next study we are planning 
to discribe the procedure of Coulomb quantisation for the Uehling potential. Note that 
our numerical algorithms based on Eqs.(\ref{total}) - (\ref{totalt}) are very 
fast and effective for accurate calculations of atomic systems. Briefly, we can say 
that the use of this closed analytical formula for the Uehling potential opens a 
new avenue in investigation of the vacuum polarisation in light and heavy atoms and 
ions and muonic atoms and ions. 

\begin{center}
    {\bf Acknowledgements}
\end{center}

It is a pleasure to acknowledge the University of Western Ontario for
financial support. One of us (AMF) wants to thank Professor Joachim Reinhardt 
(Frankfurt, Germany) for his help with Eqs.(\ref{Ueh}) and (\ref{UehA}).


\begin{thebibliography}{10}

\bibitem{Dir}P.A.M. Dirac, {\it The Principles of Quantum Mechanics.}, (Oxford
at the Clarendon Press, Oxford, UK (1930)).

\bibitem{Heis}W. Heisenbrg and H. Euler, Z. Physik {\bf 38}, 314 (1936).

\bibitem{Paul}W. Pauli and F. Villars, Rev. Mod. Phys. {\bf 21}, 434 (1949).

\bibitem{PRA1976}L. Wayne Fulleton and G.A. Rinker, Phys. Rev. A {\bf 13}, 
1283 (1976).

\bibitem{Uehl}E.A. Uehling, Phys. Rev. {\bf 48}, 55 (1935).

\bibitem{NIST}The NIST Reference on Constants, Units and Uncertainty,
(see: http://physics.nist.gov/cuu/Constants/index.html).

\bibitem{AB}A.I. Akhiezer and V.B. Beresteskii, {\it Quantum
Electrodynamics}, (4th Ed., Nauka (Science), Moscow (1981)), Chps. 4 and 5
(in Russian).

\bibitem{Grei}W. Greiner and J. Reinhardt, {\it Quantum Electrodynamics.}
(4th. Ed., Springer Verlag, Berlin, (2010)).

\bibitem{Fro1}A.M. Frolov and D.M. Wardlaw, ArXiv: 1110.3433 [nucl-th] (2011).

\bibitem{GR}I.S. Gradstein and I.M. Ryzhik, {\it Tables of Integrals, Series
and Products.}, (5th ed., Academic Press, New York, (1994)).

\bibitem{AS}{\it Handbook of Mathematical Functions.}, M. Abramowitz and
I.A. Stegun, eds. (Dover, New York, (1972)).

\bibitem{Treat}G.N. Watson, \textit{Treatise on the Bessel Functions}
(2nd. ed., McGraw-Hill, New York, Cambridge University Press (1966)).

\bibitem{Grein2}W. Pieper and W. Greiner, Nucl. Phys. A {\bf 109}, 539
(1968).

\bibitem{Plun}G. Plunien and G. Soff, Phys. Rev. A {\bf 51}, 1119 (1995).

\bibitem{WK}E.H. Wichmann and N.M. Kroll, Phys. Rev. {\bf 101}, 843 (1956).

\bibitem{Fro2}A.M. Frolov, ArXiv: 1111.2303 [math-ph] (2011).

\end{thebibliography}
\end{document}